\documentclass[aps,twocolumn,pra,superscriptaddress,amsmath,amssymb]{revtex4-2}
\usepackage{dcolumn}
\usepackage{bm}
\usepackage{amsmath}
\usepackage{mathrsfs}
\usepackage{txfonts}
\usepackage[T1]{fontenc}
\usepackage{xspace}
\usepackage{ulem}
\usepackage{comment}
\usepackage{braket}
\usepackage{lipsum}
\usepackage{mathtools}
\newcommand{\means}[1]{\langle#1\rangle}

\ifx\pdfoutput\undefined
\usepackage[dvipdfmx]{graphicx}
\usepackage[dvipdfmx]{hyperref}
\usepackage[dvipdfmx]{color}
\usepackage[dvipdfmx]{xcolor}
\else
\usepackage{graphicx}
\usepackage{hyperref}
\usepackage{color}
\usepackage{xcolor}
\usepackage[version=3]{mhchem}
\fi
\hypersetup{
        colorlinks=true,
        citecolor=blue,
        urlcolor=blue,
        linkcolor=blue
}

\begin{document}
\let\emph\textit

\title{
Magnetic-Field Effect on Excitonic Condensation Emergent in Extended Falicov-Kimball Model
}
\author{Naoya Ohta$^*$}
\author{Joji Nasu}
\affiliation{
  Department of Physics, Tohoku University, Sendai, Miyagi 980-8578, Japan
}

\date{\today}
\begin{abstract}
  We investigate the effects of magnetic fields on excitonic condensation in the extended Falicov-Kimball model, which is a spinless two-orbital Hubbard model with orbital splitting.
  In lattice systems under magnetic fields up to several tens of teslas, Zeeman effects on electron spins have been extensively studied, while the impact on orbital motion has often been considered negligible.
  However, the recent capability to generate ultra-high magnetic fields exceeding 1000~T has renewed interest in understanding their influence on ordered phases in correlated electron systems, beyond spin-related phenomena.
  To examine these effects, we incorporate a magnetic field into the extended Falicov-Kimball model by introducing the Peierls phase into the transfer integrals, thereby enabling the study of orbital motion.
  Using the Hartree-Fock approximation, we reveal a nonmonotonic response of the order parameter for excitonic condensation to increasing magnetic fields.
  At sufficiently high fields, the excitonic order parameter is suppressed, resulting in a disordered insulating state characterized by partial occupation of the two orbitals with nonzero Chern numbers.
  This state is distinct from a fully orbital-polarized configuration.
  These findings provide insights into the interplay between orbital motion and magnetic fields in multi-orbital correlated electron systems.
\end{abstract}
\maketitle

%%%%%%%%%%%
% Introduction

\section{Introduction}
\label{introduction}

Magnetic fields are widely used as external fields to alter the state of matter, similar to electric fields.
In particular, in strongly correlated electron systems, various quantum phases emerge under applied magnetic fields, which serve as crucial parameters for controlling these phases.
One of the most well-known effects of magnetic fields is the Zeeman effect, which splits energy levels according to the spin states of electrons.
By exploiting this effect, novel phenomena, such as colossal magnetoresistance~\cite{Tokura1999,Tokura462} and multiferroics in transition metal compounds~\cite{wang2003epitaxial,kimura2003magnetic,Katsura2005,Mostovoy2006,cheong2007multiferroics,khomskii2006multiferroics}, have been extensively studied in strongly correlated electron systems.
Furthermore, magnetic fields can induce exotic quantum states of matter in Mott insulators, which are regarded as quantum spin systems.
For instance, a magnetization plateau attributed to quantum many-body effects in frustrated quantum magnets~\cite{honecker2004magnetization,nishimoto2013controlling,Ishikawa2015,Suetsugu2024} and Bose-Einstein condensation of magnons induced by applied magnetic fields~\cite{Nikuni2000,giamarchi2008bose} have been both predicted and observed experimentally.

In addition to the Zeeman effect on electron spins, magnetic fields can also influence the orbital motion of electrons, resulting in Landau quantization in free-electron systems and the emergence of integer quantum Hall effects.
The Coulomb interaction between electrons can give rise to more exotic phenomena, such as fractional quantum Hall effects, due to the interplay between electron correlations and magnetic fields~\cite{Willett1987}.
This phenomenon can be understood through the Laughlin wave function, which is associated with fractional quasiparticle excitations~\cite{Laughlin1983,Fano1986,Wen1991Gapless,Aleiner1994,Hatsugai1993,Rudin1997,Gravier1998}.
These theoretical studies have primarily been conducted on two-dimensional electron gases under strong magnetic fields.

On the other hand, in lattice systems, a magnetic field, arising from spatial variations in the corresponding vector potential, induces fractal structures due to the mismatch between the periodicity of the lattice and the vector potential. 
This phenomenon is known as Hofstadter's butterfly~\cite{Hofstadter1976}.
The effects of Coulomb interactions on Hofstadter's butterfly spectra have been investigated within the Hubbard model~\cite{Doh1998,Mastropietro2019,Czajka2006,ding2022thermodynamics}, the Falicov-Kimball model, which is a spinless two-orbital Hubbard model~\cite{Tran2010,pradhan2016hofstadter}, and the $t$-$V$ model with intersite interactions~\cite{Mishra2016,Mishra2017}.
It has been reported that introducing Coulomb interactions causes Hofstadter's butterfly pattern to become smeared, modulating the band gap and bandwidth~\cite{Gudmundsson1995,Doh1998,Tran2010,wrobel2010falicov,Apalkov2014}.
Moreover, strong magnetic fields can delocalize electrons, thereby weakening the effects of electron correlations~\cite{Czajka2006,ding2022thermodynamics}.
It has also been pointed out that magnetic fields affect charge-order patterns in the Falicov-Kimball model~\cite{gruber1996falicov,gruber1997ground,yadav2017orbital,yadav2020metal,pradhan2023interplay,pandey2024electron}.
However, the effects of magnetic fields on quantum-mechanical hybridization between two orbitals induced by electron correlations remain less understood.
This may be because the magnetic field is often treated as a perturbation to electron correlations.
In lattice systems, the effects of magnetic fields with strengths up to several teslas on orbital motion are considered negligible, in stark contrast to the effects of electric fields on charge degrees of freedom.

Recently, ultra-high magnetic fields exceeding 1000~T have been successfully generated, prompting further investigation into their effects on correlated electron systems~\cite{nakamura2018record}.
By utilizing such strong magnetic fields, field-induced phase transitions to spin-polarized states have been examined in systems where spin degrees of freedom are quenched due to the formation of molecular orbitals or the full occupation of atomic orbitals.
For example, it has been reported that vanadium oxide VO$_2$, which is an insulator at low temperatures due to the dimerization of two V$^{4+}$ atoms, undergoes a metal-insulator transition to a spin-polarized metallic state when a strong magnetic field of approximately 300~T is applied~\cite{matsuda2020magnetic,matsuda2022magnetic}.
Furthermore, the perovskite cobaltite LaCoO$_3$ has been studied as a candidate material for field-induced phase transitions.
This is an archetypal compound exhibiting a spin crossover, where the Co$^{3+}$ ion in a high-spin state ($S=2$) at high temperatures changes to a low-spin state ($S=0$) at low temperatures~\cite{Tokura1998, Asai1998}.
Applying a magnetic field to the low-spin state forces the spins to align along the field direction.
Recent studies have proposed that spontaneously hybridized states of the low-spin state and the high-spin or intermediate-spin ($S=1$) state are stabilized before reaching the fully spin-polarized state under the application of an ultra-high magnetic field up to 600~T~\cite{Ikeda2016,Ikeda2020,Ikeda2023,Ikeda2024}.
These studies suggest that magnetic fields, through the Zeeman effect, can induce novel quantum states of matter in systems with quenched spin degrees of freedom.

Such ultra-high magnetic fields are expected to influence not only the spin degrees of freedom but also the orbital motion of electrons.
For instance, nearly half a century ago, it was proposed that Bose-Einstein condensation of excitons, which are pairs consisting of an electron in the conduction band and a hole in the valence band, could be induced by applying high magnetic fields to semimetallic systems~\cite{Fenton1968,Fukuyama1971,kuramoto1978electron,yoshioka1979electronic}.
Since it is triggered by Coulomb interactions between electrons and holes, excitonic condensation is one of the spontaneously symmetry-broken phases originating from electron correlations~\cite{Kaneko2025_rev}.
The overlap between the Landau subbands of the conduction and valence bands induced by a magnetic field generally decreases with increasing magnetic field strength.
As a result, the magnetic field is expected to transform a semimetal into a semiconductor and induce an excitonic insulator, which is an insulating state associated with the condensation of excitons, near the phase boundary of the semimetal-semiconductor transition.
The alloy Bi$_{1-x}$Sb$_x$, which has attracted attention as a topological insulator~\cite{Teo2008,hsieh2008topological,Zhang2009,hsieh2009observation,Taskin2009}, has been anticipated to be a candidate material for this state; however, experimental evidence for an excitonic insulator has yet to be observed.

In recent years, candidate materials for excitonic insulators, such as the quasi-one-dimensional compound Ta$_2$NiSe$_5$~\cite{Wakisaka2009,Kaneko2013,Seki2014,lu2017zero} and the transition metal dichalcogenides TiSe$_2$~\cite{Traum1978,Cercellier2007,Kogar2017} and WTe$_2$~\cite{wang2021landau,sun2022evidence,jia2022evidence}, have been proposed, with experimental observations via angle-resolved photoemission spectroscopy indicating signatures of spontaneous band hybridization~\cite{Wakisaka2009}.
This development has sparked renewed interest in the study of excitonic insulators.
Moreover, vanadium-based kagome compounds have been proposed as candidates for a time-reversal symmetry broken excitonic order~\cite{Mazza2023,scammell2023chiral,jiang2024van,Ingham2024_arxiv}.
The spontaneous hybridization of distinct spin states in LaCoO$_3$ can also be interpreted within the framework of excitonic condensation, which may represent a type of magnetically induced excitonic condensation driven by the Zeeman effect~\cite{Kunes2014condensation,Kunes2014instability,Nasu2016,Kaneko2014,Kaneko2015,Nasu2020,koga2024}.
Reflecting these recent findings, the possibility of excitonic condensation induced by magnetic effects on the orbital motion of electrons is being revisited for Bi$_{1-x}$Sb$_x$~\cite{Kinoshita2023}, offering an alternative mechanism to the magnetic field effect observed in LaCoO$_3$.
Therefore, it is necessary to explore new quantum phenomena that emerge under ultra-high magnetic fields, particularly from the perspective of spontaneously symmetry-broken phases originating from electronic correlations.

In this paper, we focus on the effects of the magnetic field not on spin degrees of freedom but on the orbital motion of electrons in correlated electron systems to clarify the stability of excitonic condensation against magnetic fields.
Specifically, we study the extended Falicov-Kimball model, which is a spinless two-orbital Hubbard model with orbital splitting corresponding to a crystalline field, on a square lattice under magnetic fields.
We introduce a magnetic field by incorporating the Peierls phase into the transfer integrals of the model.
We employ the Hartree-Fock approximation to clarify the stability of symmetry-broken phases against magnetic fields.
We perform calculations on a large cluster to address the site-dependent vector potential.
We find that the order parameter of excitonic condensation exhibits nonmonotonic behavior with respect to the magnetic field, and the excitonic condensation is suppressed and changes to the fully orbital-polarized state under the application of a magnetic field.
Similar behavior is observed in the excitonic supersolid phase, where both excitonic and orbital orders coexist; the excitonic order parameter becomes zero with an increasing magnetic field, while the orbital order parameter remains nonzero.
These results suggest that the magnetic field significantly affects the stability of excitonic condensation but has a weaker impact on orbital order.
Furthermore, when the parameter for the crystalline field is sufficiently large, the magnetic field induces a disordered insulating state, which is distinct from the fully orbital-polarized state.
In this disordered state, the two orbitals are partially occupied and possess nonzero Chern numbers with opposite signs.
This state is stabilized not by the gap opening due to electron correlations but by the gap between the Landau levels induced by the magnetic field.

This paper is organized as follows.
In the next section, we introduce the extended Falicov-Kimball model and the method used in this study.
In Sec.~\ref{sec:result}, we present the results of the magnetic-field effects on this model.
Before showing the magnetic-field dependence, the phase diagram of the extended Falicov-Kimball model in the absence of magnetic fields is presented in Sec.~\ref{sec:wo-mag}.
In Sec.~\ref{sec:mag-EC-ESS}, we present results for the magnetic-field effects on the excitonic condensation and the excitonic supersolid phases.
Section~\ref{sec:mag-DO} examines the origin of the disordered insulating state induced by magnetic fields.
In Sec.~\ref{sec:discussion}, we discuss the implications of our results and the perspectives for future studies.
Finally, Sec.~\ref{sec:summary} is devoted to the summary.

\section{Model and Method}
\label{sec:model-method}

\begin{figure*}[t]
  \begin{center}
  \includegraphics[width=1.7\columnwidth,clip]{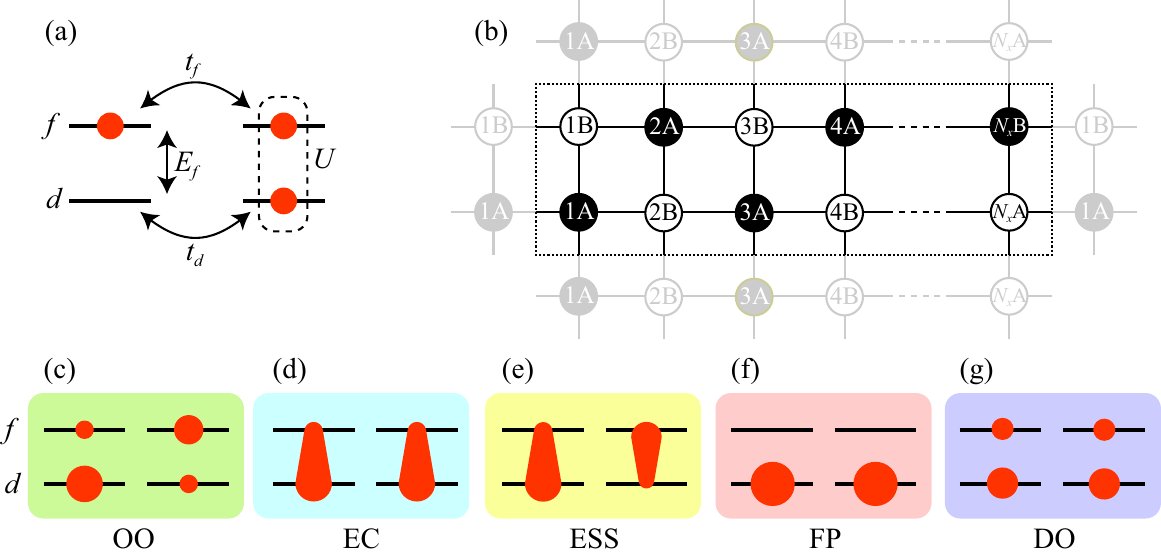}
  \caption{(Color online)
    (a) Schematic illustration of the extended Falicov-Kimball model.
    (b) Cluster including $2N_x$ sites under periodic boundary conditions for the square lattice, on which the extended Falicov-Kimball model is defined.
    Black and white circles represent the A and B sublattices, respectively.
    (c)--(g) Schematic illustrations of (c) the orbital order (OO), (d) the ferro-type excitonic condensation (EC), (e) the excitonic supersolid (ESS), (f) the fully orbital-polarized (FP) state, and (g) the disordered (DO) insulating state.
    The two lines on the left and right sides denote the two orbitals on the sites belonging to the A and B sublattices, respectively, and the size of the red objects represents the occupancy of electrons.
  }
  \label{fig:lattice}
  \end{center}
\end{figure*}

In the present study, we focus on the half-filled extended Falicov-Kimball model on a square lattice with a lattice constant $a$.
The Hamiltonian of this model is given by~\cite{Falicov1969}
\begin{align}\label{eq:Hamil}
    \mathcal{H}=-\sum_{\means{ij}}&\left(t_{ij}^d d_i^\dagger d_j
    +t_{ij}^f f_i^\dagger f_j+{\rm H.c.}\right)\notag\\
    &+U\sum_i d_i^\dagger d_i f_i^\dagger f_i
    +E_f\sum_i f_i^\dagger f_i,
\end{align}
where $d_i^\dagger$ ($f_i^\dagger$) is the creation operator for a spinless $d$ ($f$) electron at site $i$, whose position is located on the $xy$ plane and represented as $\bm{R}_i=(X_i,Y_i)$, where $X_i/a$ and $Y_i/a$ are integers.
The first term represents electron hopping between nearest-neighbor sites $\means{ij}$, with $t_{ij}^d$ and $t_{ij}^f$ denoting the transfer integrals for $d$ and $f$ electrons, respectively.
The second term describes the onsite Coulomb interaction between $d$ and $f$ electrons with magnitude $U$, and the last term represents the energy level of the $f$ electrons with $E_f$.
In the absence of magnetic fields, $t_{ij}^d$ and $t_{ij}^f$ do not depend on their site positions and are written as $t_d$ and $t_f$, respectively.
A schematic representation of the model is shown in Fig.~\ref{fig:lattice}(a).

Here, we examine the effects of a magnetic field on the extended Falicov-Kimball model, which is one of the simplest models that exhibit excitonic condensation~\cite{batista2004intermediate,brydon2008slave,schneider2008weak,zenker2010existence,phan2010spectral,phan2011excitonic,golosov2012collective,seki2012variational,farkasovsky2015influence,hamada2017excitonic,kadosawa2020finite,farkasovsky2020dmrg,farkasovsky2023hartree}.
Since this model lacks spin degrees of freedom, magnetic fields couple only with the orbital motion of electrons.
We assume that the local angular momentum is quenched for the $d$ and $f$ orbitals and is not generated through hybridization between them.
An applied magnetic field thus modulates the intersite hopping of electrons.
This effect is incorporated by introducing the Peierls phase $\theta_{ij}$ into the transfer integrals as $t_{ij}^d=t_d e^{i\theta_{ij}}$ and $t_{ij}^f=t_f e^{i\theta_{ij}}$.
Using the vector potential $\bm{A}$, the Peierls phase is expressed as
\begin{align}
    \theta_{ij}=-\frac{2\pi e}{h}\int_{\bm{R}_j}^{\bm{R}_i} \bm{A}\cdot d\bm{l},
\end{align}
where $e>0$ represents the absolute value of the electron charge and $h$ denotes the Planck constant.

We apply a static and uniform magnetic field $\bm{B}=(0,0,B)$ to the system along the $z$ direction.
This magnetic field can be represented by the spatially dependent vector potential $\bm{A}=(0,Bx,0)$, where we adopt the Landau gauge.
With this choice, the Peierls phase is given by $\theta_{ij}=1$ in the $x$ direction and $\theta_{ij}=2\pi \alpha X_i/a$ in the $y$ direction, where $\alpha=eBa^2/h$ represents the dimensionless magnetic flux.

To address the Coulomb interaction between the $d$ and $f$ electrons in Eq.~\eqref{eq:Hamil}, we adopt the Hartree-Fock approximation. This term is decoupled as $d_i^\dagger d_i f_i^\dagger f_i \to \means{d_i^\dagger d_i} f_i^\dagger f_i + d_i^\dagger d_i \means{f_i^\dagger f_i} - \means{d_i^\dagger d_i} \means{f_i^\dagger f_i} - \means{d_i^\dagger f_i} f_i^\dagger d_i - d_i^\dagger f_i \means{f_i^\dagger d_i} + \means{d_i^\dagger f_i} \means{f_i^\dagger d_i}$.
After this decoupling, the Hamiltonian is expressed as a bilinear form of fermionic operators, which we denote as $\mathcal{H}_{\rm HF}$.
Using this Hamiltonian, we compute the average $\means{c_i^\dagger c'_i}={\rm Tr}[e^{-\beta (\mathcal{H}_{\rm HF}-\mu N_e)} c_i^\dagger c'_i]/Z$, where $Z={\rm Tr}[e^{-\beta (\mathcal{H}_{\rm HF}-\mu N_e)}]$, and $c,c' = d$ or $f$.
Here, $N_e = \sum_i (d_i^\dagger d_i + f_i^\dagger f_i)$ represents the total fermion number operator.
The set of the local averages $\means{c_i^\dagger c'_i}$, which corresponds to mean fields, is determined self-consistently.
In the calculations, the chemical potential $\mu$ is chosen to satisfy the half-filling condition $\means{N_e} = N$, where $N$ denotes the number of sites.

In the absence of magnetic fields, we calculate the local average $\means{c_i^\dagger c'_i}$ by applying Fourier transformations in both the $x$ and $y$ directions under periodic boundary conditions.
On the other hand, in the presence of magnetic fields, the translational invariance by $a$ along the $x$ direction is lost because the Peierls phase depends on $X_i$.
To address this issue, we perform real-space calculations along the $x$ direction while applying Fourier transformations along the $y$ direction by introducing the wave number $k_y$.
We conduct numerical calculations within a unit cell containing $2N_x$ sites, with $N_x$ sites in the $x$ direction and $2$ sites in the $y$ direction, as shown in Fig.~\ref{fig:lattice}(b), where the mean field $\means{c_i^\dagger c'_i}$ on each site is determined independently.
We impose periodic boundary conditions as depicted in Fig.~\ref{fig:lattice}(b), and thus, the value of $\alpha$ in this system is restricted to $m/N_x$, with $m=0,1,2,\cdots,N_x-1$.
Moreover, the effect of $\alpha$ on the system is equivalent to $\alpha + \nu$ with $\nu$ being an integer, exhibiting periodicity.
Since it has been confirmed that the phases appearing in the model are independent of the sign of the magnetic field, we restrict the range of $\alpha$ to between $0$ and $1/2$ in these calculations.

To identify the states realized in the Falicov-Kimball model, we introduce the following order parameters:
\begin{align}
    \tau_{\rm F}^l=\frac{1}{N}\sum_i  \means{\bm{c}_i^\dagger \sigma^l \bm{c}_i},
\end{align}
and
\begin{align}
    \tau_{\rm AF}^l=\frac{1}{N}\sum_i (-1)^i \means{\bm{c}_i^\dagger \sigma^l \bm{c}_i},
\end{align}
where $\sigma^{l}$ with $l=x,y,z$ is the $l$ component of the Pauli matrices, $\bm{c}_i^\dagger = (d_i^\dagger, f_i^\dagger)$ is a two-component vector, and $(-1)^i$ takes the value of $+1$ ($-1$) for $i$ belonging to the A (B) sublattice as shown in Fig.~\ref{fig:lattice}(b).
Here, $\tau_{\rm AF}^z$ is an order parameter for the staggered-type orbital order presented in Fig.~\ref{fig:lattice}(c), whereas a nonzero $\tau_{\rm F}^z$ does not indicate symmetry breaking for $E_f\ne 0$. 
On the other hand, $\tau_{\rm F/AF}^x$ and $\tau_{\rm F/AF}^y$ represent spontaneous hybridization between the $d$ and $f$ orbitals since there is no inter-orbital hopping in Eq.~\eqref{eq:Hamil}.
Thus, $\tau_{\rm F}^x$ and $\tau_{\rm F}^y$ ($\tau_{\rm AF}^x$ and $\tau_{\rm AF}^y$) are regarded as order parameters for ferro-type (antiferro-type) excitonic condensation, whose phase exhibits uniform (alternating) alignment.
The ferro-type excitonic condensation is schematically shown in Fig.~\ref{fig:lattice}(d).
Note that the Hamiltonian commutes with the operator $\sum_i (f_i^\dagger f_i-d_i^\dagger d_i)$, which preserves ${\rm U(1)}$ symmetry.
This symmetry suggests degeneracy on the plane of $\tau_x$ and $\tau_y$.
We also introduce the following two quantities:
\begin{align}
    n_{\rm F}=\frac{1}{N}\sum_i  \means{d_i^\dagger d_i+f_i^\dagger f_i},
\end{align}
and
\begin{align}
    n_{\rm AF}=\frac{1}{N}\sum_i (-1)^i \means{d_i^\dagger d_i+f_i^\dagger f_i},
\end{align}
where we choose the chemical potential so as to impose $n_{\rm F}=1$ under the half-filling condition, and $n_{\rm AF}$ is the staggered-type charge order parameter.

\section{Result}
\label{sec:result}

\subsection{Phase diagram without magnetic fields}
\label{sec:wo-mag}

\begin{figure}[t]
  \begin{center}
  \includegraphics[width=\columnwidth,clip]{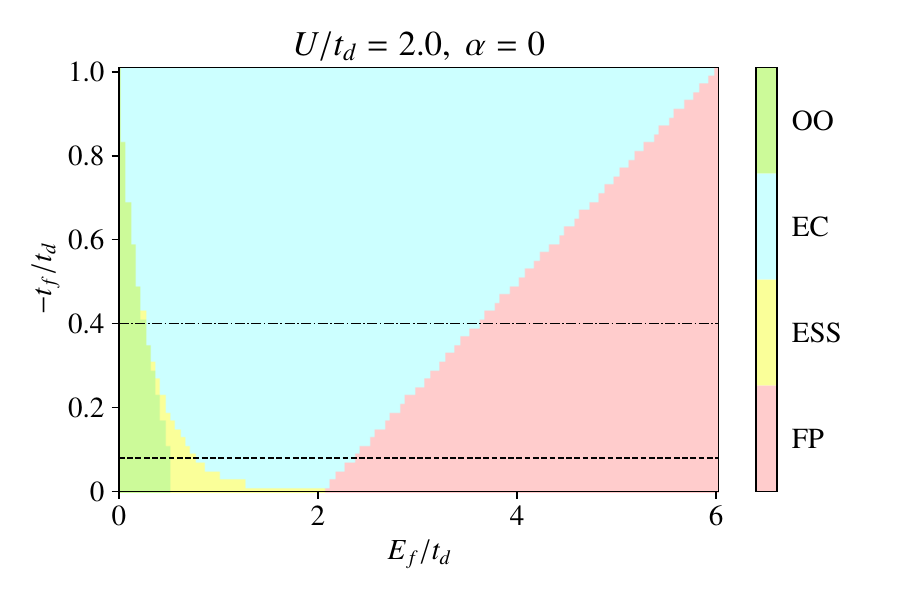}
  \caption{(Color online)
    Phase diagram of the extended Falicov-Kimball model on the $E_f$-$t_f$ plane at $U/t_d=2$ in the absence of magnetic fields, which replicates the results in Ref.~\cite{Farkasovsky2008}.
    The colored areas represent distinct phases characterized by the local states shown in Fig.~\ref{fig:lattice}(c)--\ref{fig:lattice}(f).
    The dashed line indicates the value of $t_f$ used in Fig.~\ref{fig:mag-phase}(a), and the dashed-dotted line represents the value of $t_f$ used in Fig.~\ref{fig:mag-phase}(b).
  }
  \label{fig:nonmag-phase}
  \end{center}
\end{figure}

Before presenting magnetic-field effects on the Falicov-Kimball model, we review the properties of this model in the absence of magnetic fields, as obtained by previous studies.
Figure~\ref{fig:nonmag-phase} shows the phase diagram on the $E_f$-$t_f$ plane obtained using the Hartree-Fock approximation for the two-sublattice system with $U/t_d=2$~\cite{Farkasovsky2008}.
The calculations are performed with a discretization of momenta into $128\times 128$ points.
The phase diagram consists of four distinct phases.
In the region of small $E_f$, the staggered orbital order, labeled as OO, is found and is schematically depicted in Fig.~\ref{fig:lattice}(c).
This phase is characterized by a nonzero $\tau_{\rm AF}^z$, while the excitonic order parameters $\tau^x$ and $\tau^y$ remain zero.
Conversely, in the region of large $E_f$, the fully orbital-polarized (FP) phase with $\tau_F^z=1$, shown schematically in Fig.~\ref{fig:lattice}(f), trivially appears.
Between the OO and FP phases, two distinct excitonic phases are identified.
One of these is the phase exhibiting uniform excitonic condensation, which we refer to as EC.
This phase is characterized by a nonzero $\tau_{\rm F}^x$ or $\tau_{\rm F}^y$, as shown schematically in Fig.~\ref{fig:lattice}(d).
Note that the emergence of uniform excitonic condensation is attributed to a direct gap system with $t_f/t_d<0$.
The other phase is the excitonic supersolid (ESS), where both excitonic and orbital orders coexist [see Fig.~\ref{fig:lattice}(e)].
As expected, this phase appears between the OO and EC phases and is stabilized when the amplitude of $f$-orbital hopping is small.

\begin{figure}[t]
  \begin{center}
  \includegraphics[width=\columnwidth,clip]{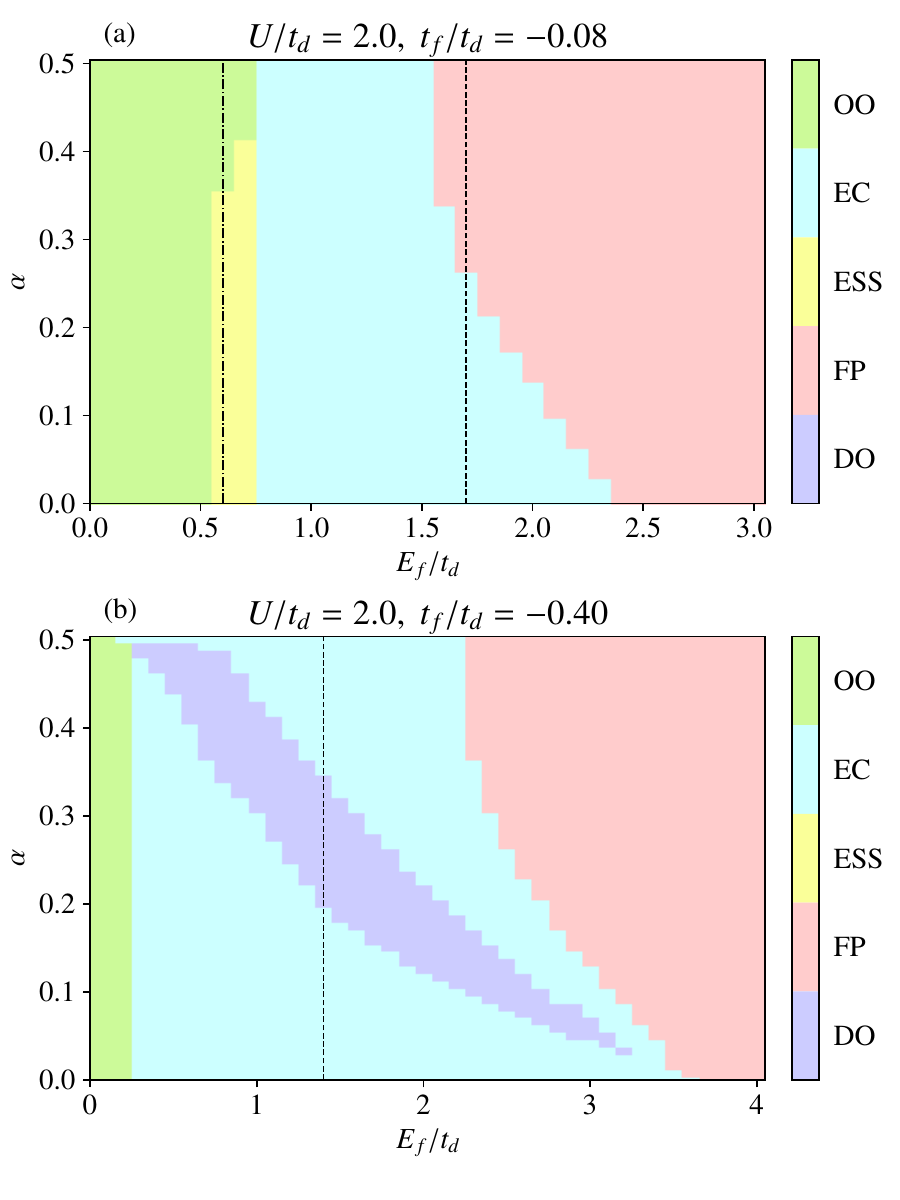}
  \caption{(Color online)
Magnetic-field phase diagram of the extended Falicov-Kimball model on the $E_f$-$\alpha$ plane at (a) $t_f/t_d=-0.08$ and (b) $t_f/t_d=-0.4$.
The dashed and dashed-dotted lines in panel (a) indicate the values of $E_f$ used in Figs.~\ref{fig:170_008} and \ref{fig:060_008}, respectively.
In panel (b), the data for $\alpha=0.5$ are obtained using calculations with a four-sublattice system.
The dashed line in panel (b) represents the value of $E_f$ used in Fig.~\ref{fig:140_040}.
}
  \label{fig:mag-phase}
  \end{center}
\end{figure}

\begin{figure}[t]
  \begin{center}
  \includegraphics[width=0.9\columnwidth,clip]{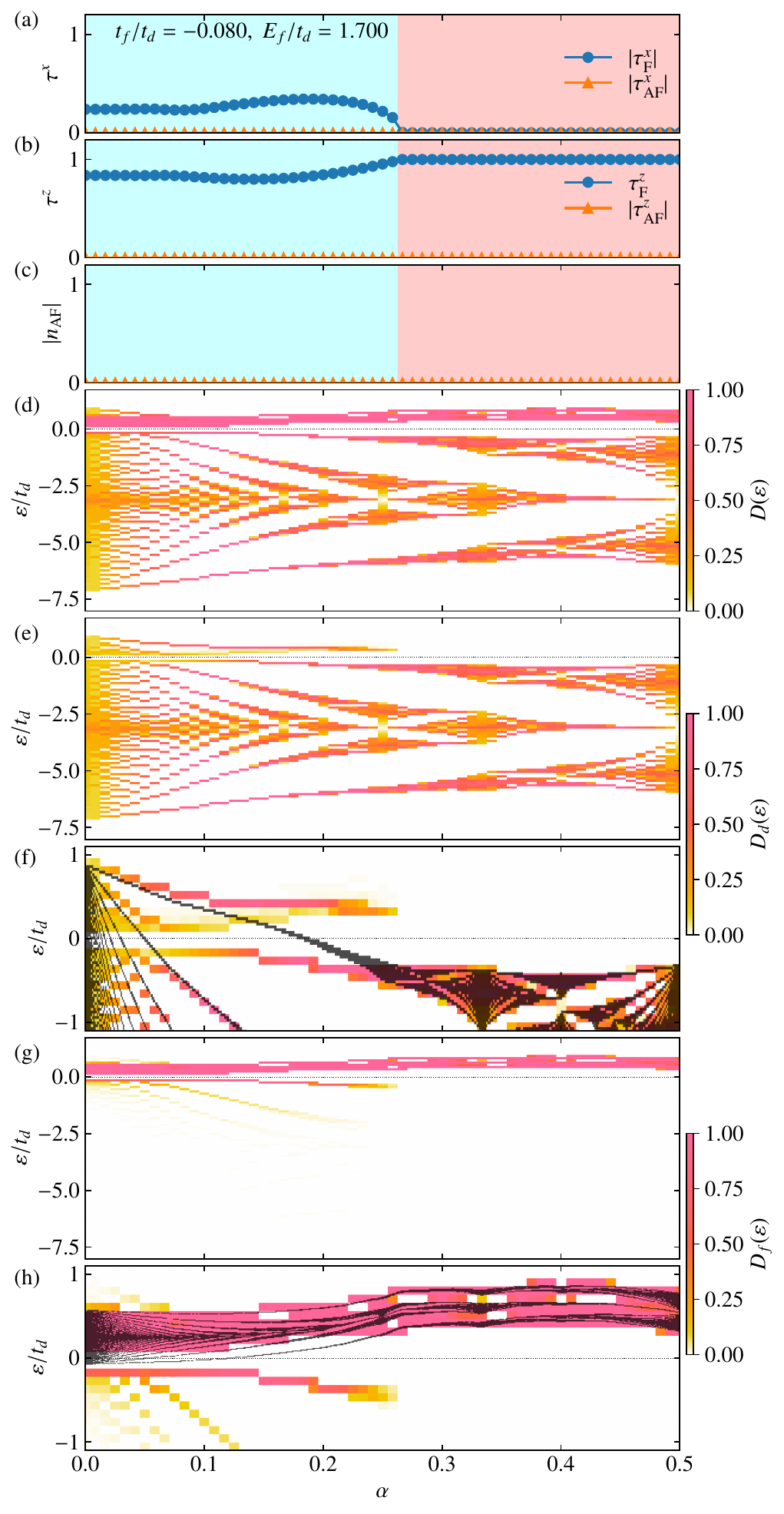}
  \caption{(Color online)
Magnetic-field dependence of (a) the excitonic order parameters $\tau_{\rm F}^x$ and $\tau_{\rm AF}^x$, (b) orbital order parameter $\tau_{\rm AF}^z$ in addition to $\tau_{\rm F}^z$, (c) charge order parameter $n_{\rm AF}$ in addition to $n_{\rm F}$, (d) DOS $D(\varepsilon)$, (e) partial DOS $D_d(\varepsilon)$ for the $d$ orbital, and (g) $D_f(\varepsilon)$ for the $f$ orbital at $t_f/t_d=-0.08$ and $E_f/t_d=1.7$.
Panels (f) and (h) display enlarged views of the DOS near the Fermi energy for the $d$ and $f$ orbitals, respectively.
In panels (d)--(f), the Fermi level is set to zero.
The noninteracting $d$- ($f$-)electron DOS is also shown in panel (f) [(h)].
  }
  \label{fig:170_008}
  \end{center}
\end{figure}

\subsection{Magnetic-field effect on EC and ESS phases}
\label{sec:mag-EC-ESS}

Here, we examine the magnetic-field effects on the Falicov-Kimball model.
The following calculations are performed for a cluster with $N_x=120$, and the Coulomb interaction is set to $U/t_d=2$, similar to the case without a magnetic field shown in Fig.~\ref{fig:nonmag-phase}.
Note that the cluster size dependence is sufficiently small, as discussed in Appendix~\ref{app:size-dependence}.
Figure~\ref{fig:mag-phase}(a) shows the magnetic-field phase diagram for $t_f/t_d=-0.08$.
We find that the EC phase becomes destabilized when the magnetic field is applied, and the region of the FP phase expands.
To clarify the origin of this behavior, we focus on the $\alpha$ dependence at $E_f/t_d=1.7$.
Figure~\ref{fig:170_008} presents the order parameters and the density of states (DOS) $D(\varepsilon)$ as functions of $\alpha$.
As demonstrated in the previous section, since the Hamiltonian possesses U(1) symmetry, we choose order parameter of excitonic condensation such that only the $\tau^x$ component is nonzero and $\tau^y$ is zero.
As shown in Fig.~\ref{fig:170_008}(a), the excitonic order parameter $\tau_{\rm F}^x$ is nonzero in the EC phase but continuously approaches zero at the critical point, indicating a transition to another phase.
In the high-field phase, Fig.~\ref{fig:170_008}(b) shows $\tau_{\rm F}^z=1$, which indicates that all electrons occupy the $d$ orbital; hence, we term this phase the fully orbital-polarized (FP) phase.
Figure~\ref{fig:170_008}(c) shows the charge order parameters.
In both the EC and FP phases, $n_{\rm AF}$ is zero, suggesting no charge disproportionation.

We also present the magnetic-field dependence of the total DOS $D(\varepsilon)$ in Fig.~\ref{fig:170_008}(d), which is defined such that $\int D(\varepsilon)d\varepsilon =2$, and the Fermi level is set to zero.
In this figure, two distinct structures are observed: a broad band structure at lower energy and a sharp, narrow band at higher energy.
To identify the origin of these structures, we present the partial DOS $D_d(\varepsilon)$ and $D_f(\varepsilon)$ for the $d$ and $f$ orbitals in Figs.~\ref{fig:170_008}(e) and \ref{fig:170_008}(g), respectively.
Note that $D_d(\varepsilon)+D_f(\varepsilon)=D(\varepsilon)$ is satisfied.
These figures clearly show that the lower energy structure originates from the $d$ orbital, while the higher energy structure originates from the $f$ orbital.
Moreover, in the region of small $\alpha$, while an energy gap exists at the Fermi energy, both $D_d(\varepsilon)$ and $D_f(\varepsilon)$ remain nonzero near this energy.
The DOS is zero at the Fermi energy, but it is finite for both $D_d(\varepsilon)$ and $D_f(\varepsilon)$ in the regions slightly above and below it.
This suggests the presence of a hybridization gap in the low-field region, corresponding to the excitonic condensation.

To observe the hybridization gap more clearly, we also present the DOS for the noninteracting $d$ and $f$ electrons. 
Enlarged views of $D_d(\varepsilon)$ and $D_f(\varepsilon)$ near the Fermi level are shown in Figs.~\ref{fig:170_008}(f) and \ref{fig:170_008}(h), respectively.
Here, the $\alpha$-dependent chemical potentials for plotting the noninteracting DOS are estimated from the center of the weight in $D_d(\varepsilon)$ and $D_f(\varepsilon)$ for a fixed $\alpha$.
We have confirmed that the overall structure of $D_d(\varepsilon)$ overlaps with the noninteracting DOS, but deviations are observed near the Fermi level in the region of the EC phase.
Due to the small magnitude of $|t_f/t_d|$, such structures are difficult to detect in the $f$ orbital; however, the weight of $D_f(\varepsilon)$ slightly below the Fermi energy is likely caused by hybridization with the $d$ orbital, because the weights of both $D_f(\varepsilon)$ and $D_d(\varepsilon)$ are nonzero around $\varepsilon/t_d= -0.1$, as shown in Figs.~\ref{fig:170_008}(f) and \ref{fig:170_008}(h).

As shown in Fig.~\ref{fig:170_008}(f), the noninteracting DOS exhibits structures composed of lines in the small $\alpha$ region, which are attributed to Landau quantization.
The second-highest Landau level intersects the Fermi level at $\alpha\simeq 0.04$, while the highest Landau level intersects it at $\alpha\simeq 0.18$, as depicted in Fig.~\ref{fig:170_008}(f).
In these regions, the spontaneous hybridization between the $d$ and $f$ orbitals is promoted by the mixing of Landau levels, leading to excitonic condensation.
Conversely, in the case where $0.04 \lesssim \alpha \lesssim 0.18$, the Fermi energy lies between the highest and second-highest Landau levels, and the excitonic order parameter is suppressed.
This observation aligns with the nonmonotonic behavior of $\tau^x$, where a dip appears around $\alpha\simeq 0.09$ in Fig.~\ref{fig:170_008}(a).

We also find that near the critical point, $\tau^x$ begins to decrease, while the gap continues to increase.
This behavior can be attributed to the reduction in bandwidth with the increase in $\alpha$, as observed in Hofstadter's butterfly pattern.
Due to the reduction in bandwidth for both $d$ and $f$ orbitals, the gap across the Fermi level can be enhanced by applying magnetic fields without spontaneous hybridization between the orbitals.
This mechanism suppresses the excitonic order parameter and transitions the EC phase into the FP phase.
The expansion of the FP phase in the phase diagram on the $E_f$-$\alpha$ plane, as shown in Fig.~\ref{fig:mag-phase}(a), can also be explained by these considerations.

\begin{figure}[t]
  \begin{center}
  \includegraphics[width=0.9\columnwidth,clip]{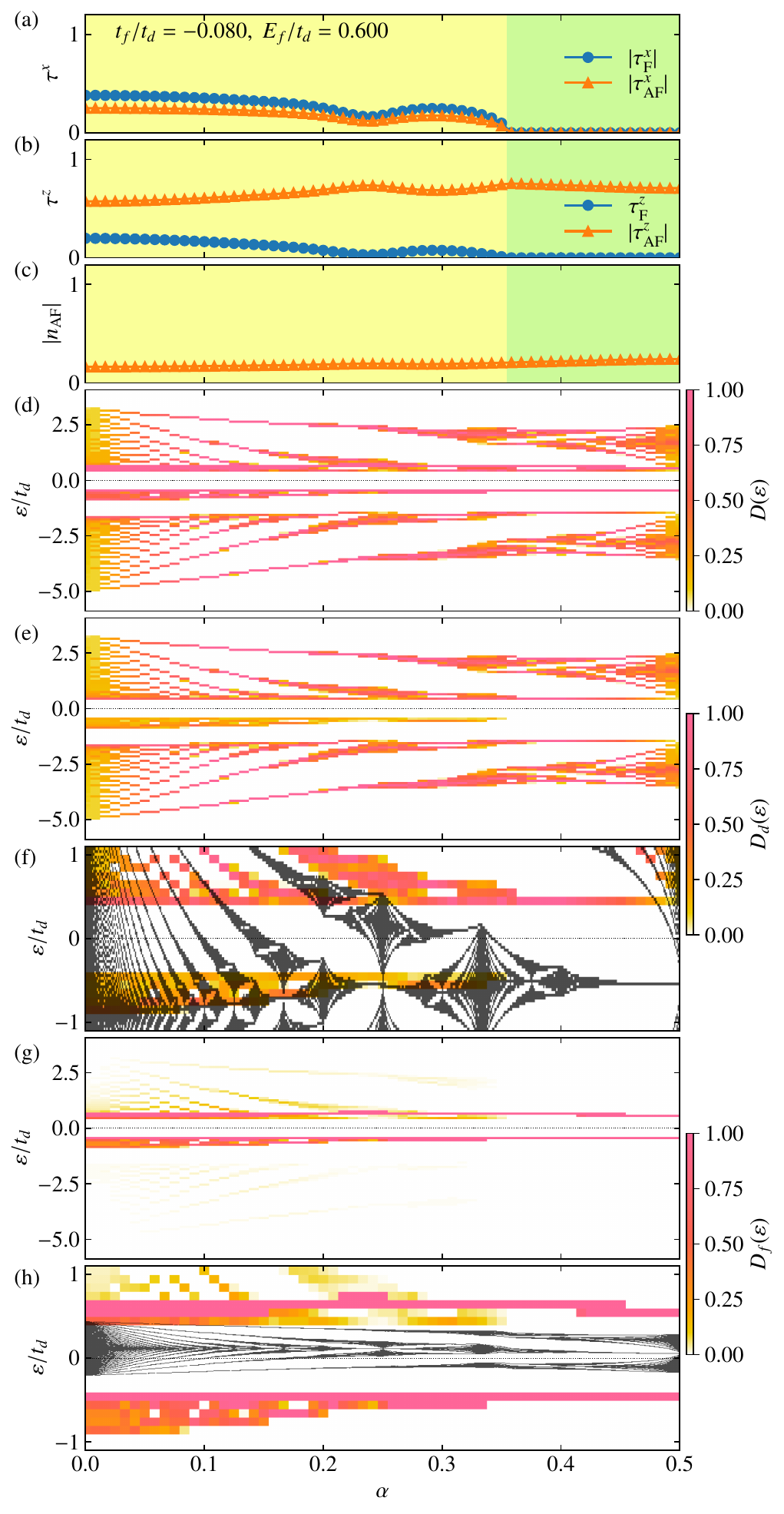}
  \caption{(Color online)
    Corresponding plot to Fig.~\ref{fig:170_008} at $t_f/t_d=-0.08$ and $E_f/t_d=0.6$.
  }
  \label{fig:060_008}
  \end{center}
\end{figure}

Next, we focus on the low-$E_f$ region, where the OO and ESS phases emerge, as shown in Fig.~\ref{fig:mag-phase}(a).
We find that the ESS phase becomes destabilized by applying the magnetic field and changes into the OO phase.
To analyze the transition between these phases in more detail, we present the $\alpha$ dependence of the order parameters at $E_f=0.6$ in Fig.~\ref{fig:060_008}.
Both ferro- and antiferro-type excitonic order parameters are nonzero in the ESS phase and exhibit similar $\alpha$ dependence, as illustrated in Fig.~\ref{fig:060_008}(a).
With a further increase in $\alpha$, these excitonic order parameters continuously approach zero at the boundary with the OO phase.
Moreover, Fig.~\ref{fig:060_008}(b) indicates that $\tau_{\rm AF}^z$ remains nonzero in both the ESS and OO phases.
These parameters exhibit nonmonotonic behavior as a function of $\alpha$.
In particular, $\tau^x$ shows a dip around $\alpha=0.24$.
Figure~\ref{fig:060_008}(c) illustrates the $\alpha$ dependence of $n_{\rm AF}$, which signifies the presence of charge disproportionation in both the ESS and OO phases.
This behavior arises from the inequivalence of the two orbitals induced by the nonzero $E_f$, as previously discussed in the two-orbital Hubbard model with a crystalline field~\cite{koga2024}.

To understand the origin of the nonmonotonic behavior of the excitonic order parameters, we examine the $\alpha$ dependence of the DOS.
As shown in Fig.~\ref{fig:060_008}(d), there are two band gaps at $\alpha=0$.
In the sharp peaks around $\varepsilon/t_d=\pm 0.6$, both $D_d(\varepsilon)$ and $D_f(\varepsilon)$ are nonzero, as shown in Fig.~\ref{fig:060_008}(e) and \ref{fig:060_008}(g).
As presented in Fig.~\ref{fig:060_008}(h), the noninteracting $f$-electron DOS is located between the two structures of $D_f(\varepsilon)$.
This suggests that the gap opening arising from the Coulomb interaction is not primarily due to the spontaneous hybridization between the $d$ and $f$ orbitals but instead attributed to the formation of a staggered-type orbital ordering.
The above interpretation is also supported by the fact that such a feature in $D_f(\varepsilon)$ is maintained in the region of the OO phase in addition to the ESS phase.
The difference between the ESS and OO phases is the presence of excitonic condensation.
We focus on the isolated band around $\varepsilon/t_d=-0.6$ in $D_d(\varepsilon)$, as shown in Fig.~\ref{fig:060_008}(e) and its enlarged view near the Fermi energy, Fig.~\ref{fig:060_008}(f).
In the ESS phase, $D_f(\varepsilon)$ is also nonzero in this energy region, as shown in Fig.~\ref{fig:060_008}(h), suggesting the presence of spontaneous hybridization between the $d$ and $f$ orbitals.
As $\alpha$ increases, the $d$ orbital contribution to the isolated band around $\varepsilon/t_d=-0.6$ weakens, with the weight diminishing around $\alpha=0.24$ and vanishing near $\alpha=0.36$, which resembles the nonmonotonic behavior of $\tau^x$.

Here, we focus on the low-energy region of $D_d(\varepsilon)$, as shown in Figs.~\ref{fig:060_008}(f).
Under the application of a magnetic field, Landau quantization occurs.
We observe that the suppression of $D_d(\varepsilon)$ for the isolated band with $\varepsilon/t_d \sim -0.6$ around $\alpha=0.23$ occurs in the region where the Fermi energy lies within the gap between the second- and third-highest Landau levels for the noninteracting $d$-electron DOS.
Additionally, the excitonic order appears to vanish in the region where the Fermi energy lies within the gap between the highest and second-highest Landau levels for the noninteracting $d$-electron DOS.
These findings suggest that excitonic condensation is promoted by the mixing of Landau levels between the $d$ and $f$ orbitals and suppressed by the gap opening between Landau levels, resulting in the nonmonotonic behavior of the excitonic order parameter.
Similar phenomena have been discussed as the origin of the quantum oscillations observed in candidate materials for topological Kondo insulators~\cite{Knolle2015Quantum,Peters2019} and WTe$_2$, which is a candidate for excitonic insulators~\cite{Lee2021quantum,Lee2021quantum-thermally,Andrew2022,Allocca2024}.

On the other hand, the antiferro-type orbital order is not destabilized by applying a magnetic field, in contrast to the excitonic order.
This characteristic can be observed in Fig.~\ref{fig:mag-phase}(a), where the phase boundary on the left side of the EC phase remains almost independent of $\alpha$.
These results suggest that the orbital order is unaffected by the magnetic field, likely because the local orbital angular momentum is not taken into account.
When considering orbitals with atomic orbital angular momentum, it is anticipated that the phase boundary may depend on the magnetic field due to the Zeeman effect.
In this scenario, as the Zeeman effect mixes real orbitals, such a magnetic field might suppress the orbital order~\cite{koga2024}.

\subsection{Magnetic-field induced disorder phase}
\label{sec:mag-DO}

\begin{figure}[t]
  \begin{center}
  \includegraphics[width=0.9\columnwidth,clip]{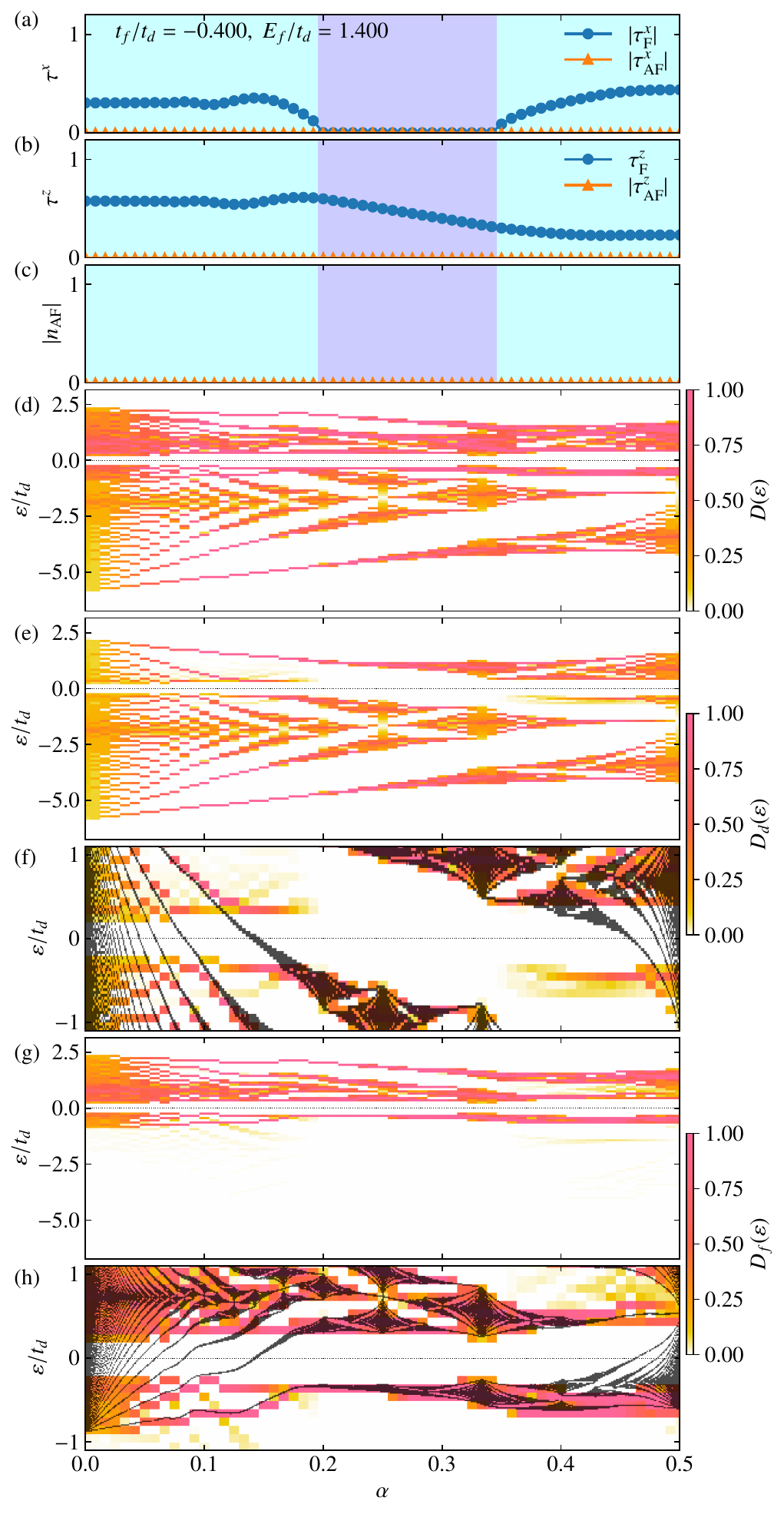}
  \caption{(Color online)
    Corresponding plot to Fig.~\ref{fig:170_008} at $t_f/t_d=-0.4$ and $E_f/t_d=1.4$.
  }
  \label{fig:140_040}
  \end{center}
\end{figure}

In the previous section, we examined the case with small $|t_f|$ compared to $t_d$, where the ESS phase appears between the OO and EC phases.
In this section, we focus on the case with a larger $|t_f|$.
As shown in Fig.~\ref{fig:nonmag-phase}, increasing $|t_f|$ suppresses the ESS phase and stabilizes the EC phase due to the significant overlap between the two orbitals.
Here, we set the parameter as $t_f/t_d = -0.4$.
The magnetic-field phase diagram obtained in the present calculations is given in Fig.~\ref{fig:mag-phase}(b).
We identify a disordered (DO) phase, where the indicators of the symmetry breakings corresponding to excitonic condensation and orbital and charge orders and are zero.
Such a disordered phase does not appear in the cases with $\alpha=0$ and $\alpha=0.5$.
Note that in the latter case, known as the $\pi$-flux state, linear dispersions appear with point nodes, hindering the convergence of physical quantities for $N_x$.
In Fig.~\ref{fig:mag-phase}(b), to reduce the slow convergence, calculations with $\alpha=0.5$ are performed separately in the four-sublattice system where the wavenumber sum was taken as $128 \times 128$ in both the $x$ and $y$ directions.
The phase diagram at $\alpha=0.5$ on the $E_f$-$t_f$ plane, similar to Fig.~\ref{fig:nonmag-phase}, is provided in Appendix~\ref{app:pi-flux} and Fig.~\ref{fig:05mag-phase}.
Importantly, the disordered phase found in Fig.~\ref{fig:mag-phase}(b) does not appear in this phase diagram at $\alpha=0.5$.

To clarify the origin of the DO phase, we examine the magnetic-field dependence of the order parameters and DOS.
Figure~\ref{fig:140_040}(a) displays the excitonic order parameters at $E_f=1.4$, as indicated by the dashed line in Fig.~\ref{fig:mag-phase}(b).
As illustrated in this figure, $\tau_{\rm AF}^x$ is always zero.
In contrast, $\tau_{\rm F}^x$ oscillates as $\alpha$ increases in the small $\alpha$ region and continuously approaches zero.
In the phase where $\tau_{\rm F}^x=0$, both $\tau^z_{\rm AF}$ and $n_{\rm AF}$ are also zero, as depicted in Figs.~\ref{fig:140_040}(b) and \ref{fig:140_040}(c), indicating the absence of orbital and charge orders in addition to excitonic condensation.
We observe a revival of the excitonic order parameter $\tau_{\rm F}^x$ above $\alpha\simeq 0.35$.
As presented in Fig.~\ref{fig:140_040}(d), a gap persists at the Fermi level for all values of $\alpha$, indicating that the DO phase is insulating.

The oscillating behavior of $\tau_{\rm F}^x$ in the EC phase is attributed to the same cause as the nonmonotonic $\alpha$ dependence of $\tau_{\rm F}^x$, which arises from Landau quantization as discussed in Sec.~\ref{sec:mag-EC-ESS}.
Figures~\ref{fig:140_040}(e) and \ref{fig:140_040}(g) show the DOS for the $d$ and $f$ orbitals, respectively, while Figs.~\ref{fig:140_040}(f) and \ref{fig:140_040}(h) provide enlarged views of the regions around the Fermi level.
The dip of $\tau_{\rm F}^x$ around $\alpha \simeq 0.1$ is related to the Fermi energy being located between the second-highest (second-lowest) and third-highest (third-lowest) Landau levels for the noninteracting $d$- ($f$-)electron DOS.
In addition, the broad peak structure of $\tau_{\rm F}^x$ around $\alpha \simeq 0.15$ is attributed to the second-highest (second-lowest) Landau level crossing the Fermi level, which enhances excitonic condensation due to spontaneous hybridization between the two orbitals.
Further increases in $\alpha$ result in the narrowing of the bandwidths, causing the Fermi energy to lie between the highest (lowest) and second-highest (second-lowest) Landau levels for the noninteracting $d$- ($f$-)electron DOS.
Since these gaps are largest for the corresponding orbitals, they strongly suppress excitonic condensation, resulting in the realization of the DO phase.
In the DO phase, all electrons (holes) in the $f$ ($d$) orbital occupy the lowest (highest) Landau level, corresponding to quantum-limit states.
We have confirmed that the Chern numbers of the $d$ and $f$ orbitals are $\pm 1$ with opposite signs.

As $\alpha$ increases further, the excitonic order parameter is revived at $\alpha \sim 0.35$, as shown in Fig.~\ref{fig:140_040}(a).
At this strength of the magnetic field, the Landau levels become smeared due to the fractal structure inherent to Hofstadter's butterfly bands, as presented in Figs.~\ref{fig:140_040}(e)--\ref{fig:140_040}(h).
The increase in the bandwidths enhances the spontaneous mixing between the $d$ and $f$ orbitals.
This mechanism stabilizes excitonic condensation, which is considered to be the origin of the reentrant behavior of the EC phase.
Therefore, we conclude that the reentrant behavior of the EC phase is a characteristic feature of lattice systems.

\section{Discussion}
\label{sec:discussion}

Here, we discuss the possibilities of realizing the magnetic-field-induced quantum phenomena obtained in the present study, as well as perspectives for future research.
We can estimate the magnitude of the magnetic fields with $\alpha=1$, corresponding to $h/ea^2\sim 10^4~{\rm T}$, when the area of a unit cell perpendicular to an applied magnetic field is assumed to be $a^2\sim 10~\AA$, which is a typical scale for transition-metal compounds, although our study does not focus on specific materials.
As presented in Fig.~\ref{fig:mag-phase}, the DO phase and the reentrant behavior of the EC phase are observed up to $\alpha\sim 0.1$ in the case of a larger $E_f$ in the extended Falicov-Kimball model.
Thus, we expect that the ultra-high magnetic field of $10^3~{\rm T}$, which is currently achievable experimentally, can facilitate the examination of magnetic-field-induced quantum phenomena originating from Landau-level formations and Hofstadter's butterfly structure characteristic of lattice systems.

In the present calculations, Coulomb interactions are treated within the HF approximation.
In the case of strong Coulomb interactions, quantum many-body effects are expected to play a crucial role.
To incorporate these effects, real-space dynamical mean-field theory can be applied. This approach has been employed to investigate Mott transitions in the Falicov-Kimball model under strong magnetic fields~\cite{Tran2010} and quantum oscillations in Kondo insulators~\cite{Peters2019}.
Furthermore, considering other magnetic-field effects, such as the Zeeman effect, is essential when spin degrees of freedom are present, although these are neglected in the current study.
These aspects represent an important future direction for the study of magnetic-field-induced quantum phenomena in correlated electron systems.

In addition to neglecting spin degrees of freedom, we have simplified the lattice structure to a two-dimensional square lattice.
In real materials, the lattice structure is more complicated, and the interplay between the lattice structure and magnetic fields can give rise to various quantum phenomena.
Effects arising from such lattice structures and the complexity of hopping can be incorporated as a first step by introducing diagonal hopping into the simplest square lattice considered in this study, which corresponds to considering a connection to a triangular lattice, where geometrical frustration effects on orbital or spin orderings can be expected.
In this situation, larger magnetic unit cells are required to capture ordered states with long periodicity.
This challenge may be overcome by selecting an appropriate Peierls phase to reduce the magnetic unit cell~\cite{Cresti2021}.

We also note that strong magnetic fields can induce effects on electronic systems beyond those introduced through the incorporation of Peierls phases.
For example, a method to incorporate both the Zeeman and Paschen-Back effects has been proposed by developing a nonperturbative theory for addressing magnetic fields~\cite{Higuchi2015,Higuchi2018}.
Moreover, it has been pointed out that the spatial dependence of the vector potential affects the orthogonality of Wannier functions, which can also influence electronic properties beyond contributions from Peierls phases~\cite{Matsuura2016}.
These effects may be crucial for understanding the electronic properties of correlated electron systems under strong magnetic fields, and they should be considered in future studies.

\section{Summary}
\label{sec:summary}

In summary, we have investigated the magnetic-field effects on the Falicov-Kimball model, which describes spinless electrons in two orbitals with onsite Coulomb interactions, focusing on excitonic condensation in addition to orbital and charge orders.
To address the magnetic-field effects on this model, we incorporated the Peierls phase into the hopping terms in the Hamiltonian.
Using the Hartree-Fock approximation, we examined the ordered phases in the magnetic phase diagram.
We found an oscillating behavior of the excitonic order parameter when applying a magnetic field.
This oscillation is attributed to Landau quantization, which facilitates excitonic condensation when the Landau levels cross the Fermi energy, resulting in spontaneous hybridization between the two orbitals due to the Coulomb interaction.
Further increases in the magnetic field suppress the excitonic order, leading to the emergence of a disordered insulating state.
This disordered state is characterized by the occupation of the lowest (highest) Landau levels of the electron (hole) band, corresponding to the quantum-limit states.
In this state, each orbital possesses nonzero Chern numbers with opposite signs, which is distinct from a fully orbital-polarized state.
In the higher magnetic field region, the excitonic order parameter reemerges, indicating the reentrant behavior of excitonic condensation.
We revealed that this behavior is attributed to the fractal structure of Hofstadter's butterfly bands, which are intrinsic to lattice systems under magnetic fields.
We also demonstrated that, although the magnetic field suppresses excitonic condensation, it does not significantly affect the stability of orbital orders, as shown by examining the effects of the magnetic field on an excitonic supersolid phase.

\begin{acknowledgments}
The authors thank M.~Koshino, Y.~Tada, A.~Yamada, and A.~Ono for fruitful discussions.
Parts of the numerical calculations were performed in the supercomputing systems in ISSP, the University of Tokyo.
This work was supported by Grant-in-Aid for Scientific Research from
JSPS, KAKENHI Grant Nos.~JP19K03742, JP20H00122, JP20K14394, JP23K13052, and JP23K25805.
N.O. acknowledges support from GP-Spin at Tohoku University.
\end{acknowledgments}

\appendix

\section{Cluster size dependence}
\label{app:size-dependence}

\begin{figure}[t]
  \begin{center}
  \includegraphics[width=\columnwidth,clip]{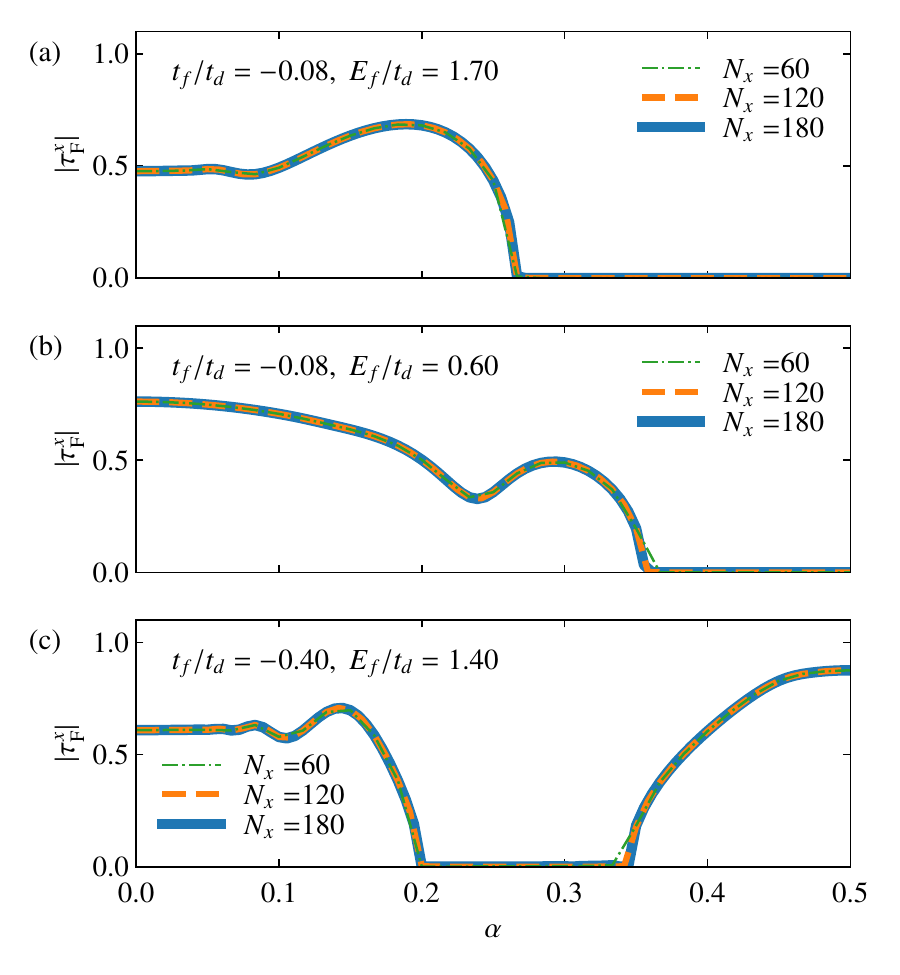}
  \caption{(Color online)
    Size dependence of the order parameter $|\tau_{\rm F}^x|$ at (a) $t_f/t_d=-0.08$ and $E_f/t_d=1.7$ corresponding to Fig.~\ref{fig:170_008}, (b) $t_f/t_d=-0.08$ and  $E_f/t_d=0.6$ corresponding to Fig.~\ref{fig:060_008}, and (c) $t_f/t_d=-0.14$ and $E_f/t_d=0.4$ corresponding to Fig.~\ref{fig:140_040}.
  }
  \label{fig:size}
  \end{center}
\end{figure}

\begin{figure}[t]
  \begin{center}
  \includegraphics[width=\columnwidth,clip]{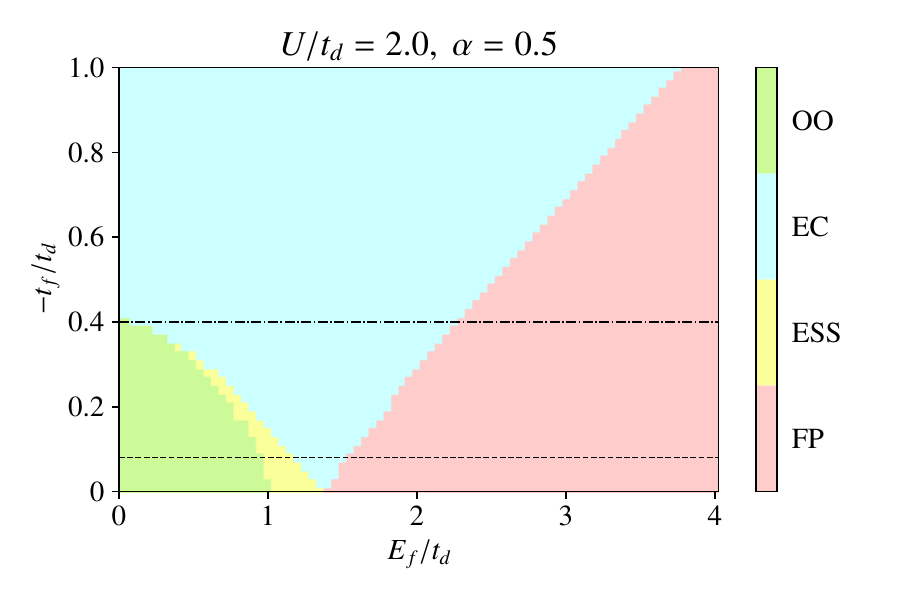}
  \caption{(Color online)
Phase diagram of the extended Falicov-Kimball model on the $E_f$-$t_f$ plane at $\alpha=0.5$.
The dashed line indicates the value of $t_f$ used in Fig.~\ref{fig:mag-phase}(a), and the dashed-dotted line represents the value of $t_f$ used in Fig.~\ref{fig:mag-phase}(b).
  }
  \label{fig:05mag-phase}
  \end{center}
\end{figure}

In this appendix, we examine the cluster size dependence of the order parameters.
Figure~\ref{fig:size} shows the $\alpha$ dependence of the order parameter $|\tau_{\rm F}^x|$ for different cluster sizes under the parameters analyzed in Figs.~\ref{fig:170_008}, \ref{fig:060_008}, and \ref{fig:140_040}.
The order parameter remains nearly unchanged for the cluster sizes presented in Fig.~\ref{fig:size}, suggesting that calculations with $N_x=120$ are sufficiently converged.

\section{Phase diagram for $\pi$-flux state}
\label{app:pi-flux}

In this appendix, we discuss the properties of the Falicov-Kimball model at $\alpha=0.5$, corresponding to the $\pi$-flux state.
We perform four-sublattice calculations, where the number of discretized momenta is $128\times 128$.
Figure~\ref{fig:05mag-phase} shows the phase diagram on the $E_f$-$t_f$ plane at $\alpha=0.5$.
The overall structure is similar to that without a magnetic field, as presented in Fig.~\ref{fig:nonmag-phase}, but the regions of the OO and ESS phases are significantly reduced.
It is noteworthy that the disordered phase does not appear even at $\alpha=0.5$.

\noindent
$^*$\texttt{naoya.ohta.p3@dc.tohoku.ac.jp}

\bibliography{./refs}
\end{document}